\documentclass[preprint]{aastex}
\pdfoutput=1
\usepackage{xspace}
\begin{document}
\newcommand{\um}{$\mu$m\xspace}
\title{Dust Production and Particle Acceleration in Supernova 1987A Revealed with ALMA}
\author{R. Indebetouw\altaffilmark{1,2},
M. Matsuura\altaffilmark{3},
E. Dwek\altaffilmark{4},
G. Zanardo\altaffilmark{5},
M.J. Barlow\altaffilmark{3},
M. Baes\altaffilmark{6},
P. Bouchet\altaffilmark{7},
D.N. Burrows\altaffilmark{8},
R. Chevalier\altaffilmark{1},
G.C. Clayton\altaffilmark{9},
C. Fransson\altaffilmark{10},
B. Gaensler\altaffilmark{11,12},
R. Kirshner\altaffilmark{13},
M. Laki\'cevi\'c\altaffilmark{14},
K.S. Long\altaffilmark{15},
P. Lundqvist\altaffilmark{10},
I. Mart\'i-Vidal\altaffilmark{16},
J. Marcaide\altaffilmark{17},
R. McCray\altaffilmark{18},
M. Meixner\altaffilmark{15,19},
C.-Y. Ng\altaffilmark{20},
S. Park\altaffilmark{21},
G. Sonneborn\altaffilmark{15},
L. Staveley-Smith\altaffilmark{5,11},
C. Vlahakis\altaffilmark{22},
J. van Loon\altaffilmark{14}}
\altaffiltext{1}{Department of Astronomy, University of Virginia, PO Box 400325, Charlottesville, VA, 22904, USA; remy@virginia.edu}
\altaffiltext{2}{National Radio Astronomy Observatory, 520 Edgemont Rd, Charlottesville, VA, 22903, USA}
\altaffiltext{3}{Department of Physics and Astronomy, University College London, Gower Street, London WC1E 6BT, UK}
\altaffiltext{4}{NASA Goddard Space Flight Center, 8800 Greenbelt Road, Greenbelt, MD 20771, USA}
\altaffiltext{5}{International Centre for Radio Astronomy Research (ICRAR), University of Western Australia, Crawley, WA 6009, Australia}
\altaffiltext{6}{Sterrenkundig Observatorium, Universiteit Gent, Krijgslaan 281 S9, B-9000 Gent, Belgium}
\altaffiltext{7}{CEA-Saclay, 91191 Gif-sur-Yvette, France}
\altaffiltext{8}{Department of Astronomy \& Astrophysics, The Pennsylvania State University, University Park, PA 16802, USA}
\altaffiltext{9}{Department of Physics and Astronomy, Louisiana State University, Baton Rouge, LA 70803}
\altaffiltext{10}{Department of Astronomy and the Oskar Klein Centre, Stockholm University, AlbaNova, SE-106 91 Stockholm, Sweden}
\altaffiltext{11}{Australian Research Council Centre of Excellence for All-sky Astrophysics (CAASTRO)}
\altaffiltext{12}{Sydney Institute for Astronomy, School of Physics, The University of Sydney, NSW 2006, Australia}
\altaffiltext{13}{Harvard-Smithsonian Center for Astrophysics, 60 Garden St., Cambridge MA 02138, USA}
\altaffiltext{14}{Lennard-Jones Laboratories, Keele University, ST5 5BG, UK}
\altaffiltext{15}{Space Telescope Science Institute, 3700 San Martin Drive, Baltimore, MD 21218, USA}
\altaffiltext{16}{Department of Earth and Space Sciences, Chalmers University of Technology, Onsala Space Observatory, SE-43992 Onsala, Sweden}
\altaffiltext{17}{Universidad de Valencia, C/Dr. Moliner 50, E-46100 Burjassot, Spain}
\altaffiltext{18}{Department of Astrophysical and Planetary Sciences, University of Colorado at Boulder, UCB 391, Boulder, CO 80309, USA}
\altaffiltext{19}{ The Johns Hopkins University, Department of Physics and
Astronomy, 366 Bloomberg Center, 3400 N. Charles Street, Baltimore, MD
21218, USA}
\altaffiltext{20}{Department of Physics, The University of Hong Kong, Pokfulam Road, Hong Kong}
\altaffiltext{21}{Department of Physics, University of Texas at Arlington, 108 Science Hall, Box 19059, Arlington, TX 76019, USA}
\altaffiltext{22}{Joint ALMA Observatory/European Southern Observatory, Alonso de Cordova 3107, Vitacura, Santiago, Chile}

\begin{abstract} Supernova (SN) explosions are crucial engines driving the
  evolution of galaxies by shock heating gas, increasing the
  metallicity, creating dust, and accelerating energetic particles.
  In 2012 we used the Atacama Large Millimeter/Submillimeter Array to
  observe SN\ 1987A, one of the best-observed supernovae since the
  invention of the telescope.  We present spatially resolved images at
  450$\,$\um, 870$\,$\um, 1.4$\,$mm, and 2.8$\,$mm, an important transition
  wavelength range.  Longer wavelength emission is dominated by
  synchrotron radiation from shock-accelerated particles, shorter
  wavelengths by emission from the largest mass of dust measured in a
  supernova remnant ($>$0.2\ M$_\odot$).  For the first time we show
  unambiguously that this dust has formed in the inner ejecta (the
  cold remnants of the exploded star's core).  The dust emission is
  concentrated to the center of the remnant, so the dust has not yet
  been affected by the shocks. If a significant fraction survives, and
  if SN\ 1987A is typical, supernovae are important cosmological dust
  producers.  
\end{abstract}

\keywords{supernovae: individual (1987A) --- ISM: supernova remnants --- galaxies: ISM --- Magellanic Clouds}

\section{Introduction}

Supernova (SN) 1987A in the Large Magellanic Cloud was the closest
supernova explosion to Earth (50kpc) observed since Kepler's SN1604AD,
making it a unique target to study supernova and supernova remnant
physics.  Supernovae are thought to be one of the most important
sources of dust in the universe.
Stars synthesize heavy
elements, and after they explode, rapid expansion and radiation cause
the ejected gas to cool rapidly, allowing refractory elements to
condense into dust grains. Dust masses have been measured in over
twenty supernovae and supernova remnants \citep{gall11}; most contain
10$^{-6}$-10$^{-3}\,$M$_\odot$ of warm dust, far below predictions of
0.1-1.0$\,$M$_\odot$ \citep{todini01,nozawa03,cherchneff10}.  One of
the few exceptions is SN 1987A -- far infrared photometry (100-350\um)
with the Herschel Space Observatory in 2011 implied
0.4-0.7$\,$M$_\odot$ of dust at 20-26$\,$K \citep{matsuura11}.
Measured 24 years after the explosion, that dust mass is significantly
higher than the 10$^{-4}\,$M$_\odot$ reported from infrared
measurements 1-2 years after the explosion
\citep{wooden93,bouchet04}. While the Herschel result supports
theoretical models of significant dust production, the discrepancy
with prior results caused many to question whether all of the dust
detected in 2011 was in fact produced in the supernova.  SN$\,$1987A's
progenitor star could have created a massive dust shell during its Red
Supergiant phase.  Mid-infrared (MIR) photometry of SN$\;$1987A in 2003
\citep{bouchet04} and spectroscopy in 2005 \citep{dwek10} only found
10$^{-5}\,$M$_\odot$ of warm progenitor dust.  However, observations of Galactic
evolved stars with masses bracketing that of SN$\,$1987A's progenitor
reveal between 0.01 and 0.4$\,$M$_\odot$ of dust (the Egg Nebula, 
\citet{jura00} and Eta Carina,  \citet{gomez10}), so if only a minor warm component of the progenitor dust was detected in the MIR, a large cold 
progenitor dust mass detected for the first time by Herschel is
plausible. Herschel had insufficient spatial resolution to distinguish
between the ejecta, shocked progenitor wind, and nearby interstellar
material.

Resolved observations of SN$\,$1987A in the submillimeter regime are
required to establish the location of the emission measured by
Herschel.  Multi-wavelength resolved images are required to determine
its physical origin, and such images can be used to simultaneously
study shock particle acceleration.  Twenty-five years after the
explosion, SN$\,$1987A's blast wave had propagated to a radius of
7$\times$10$^{17}\,$cm (0.9\arcsec on the sky), shocking material lost
from the progenitor star \citep{crotts00}.  Electrons accelerated by
the shocks produce a bright shell or torus of synchrotron radiation at
centimeter wavelengths \citep{potter09,ng13}. The shocks are
interacting with a dense equatorial ring tilted 43 degrees from the
line-of-sight \citep{tziam11}, 
dominating the emission at Xray, 
optical to thermal infrared ($<$30$\,$\um) wavelengths
\citep{bouchet04,dwek10}.  Interior to the shock and ring are the
remnants of the star's metal-rich core, referred to hereafter as the
inner ejecta.  A supernova explosion can also leave behind a black
hole or a neutron star. The observed neutrino burst implies that a
neutron star must have at least temporarily formed in SN$\,$1987A
\citep{mccray93}, but it has yet to be detected.  If the
neutron star energizes a pulsar wind nebula (PWN), its emission would
most likely have a spectrum with power-law spectral index of -0.3 to 0
\citep{gaensler06}, flatter than synchrotron emission from the shock,
perhaps detectable at millimeter wavelengths or even contributing to
the Herschel far-infrared emission.  Previous observations at 10$\,$mm
\citep{potter09}, 6.8$\,$mm \citep{zanardo13}, and 3.2$\,$mm
\citep{laki-atca} tentatively detected such excess emission at the
remnant's center, with large uncertainties from image deconvolution;
high spatial resolution images from millimeter to submillimeter
wavelengths are essential to disentangle the emission mechanisms.

\section{Observations}
ALMA observations are executed according to the quality of the
weather, and in general the observing requirements are more stringent
the higher the frequency of the observation.  SN$\,$1987A was observed
during 2012 multiple times with $\sim$20 antennas in configurations
containing baselines between $\sim$20$\,$m and $\sim$400$\,$m.  Bands
3,6,7, and 9 (2.8$\,$mm, 1.4$\,$mm, 870$\,$\um, and 450$\,$\um) were
observed in approximately chronological order. Band 3:
A002/X3c5ee0/X24b (5 Apr), A002/X3c7a84/X1c (6 Apr); Band 6:
A002/X3c8f66/X352 (7 Apr), A002/X45f1dd/Xd13 (15 Jul),
A002/X494155/X8be (10 Aug); Band 7: A002/X45e2af/X458 (14 Jul),
A002/X4ae797/X776 (24 Aug); Band 9: A002/X4afc39/X8ce (25 Aug),
A002/X4b29af/Xd21 (27 Aug), A002/X535168/X796 (05 Nov).
%
Each dataset was calibrated individually with Common Astronomy
Software Applications (CASA; \url{casa.nrao.edu}).  All data in a
given wavelength band were then combined for imaging and deconvolution
with the ``clean'' algorithm.  Only the channels free of bright CO and SiO emission were used (212.56-213.59$\;$GHz and 100.07-103.91$\;$GHz in bands 6 and 3).
In synthesis imaging the weighting as a
function of baseline length can be adjusted, to achieve finer spatial
resolution at the expense of somewhat higher noise per beam.  Briggs
weighting was used with ``robust'' parameters and resulting Gaussian
restoring beams as noted in the Figures.  ATCA images were
deconvolved using the maximum entropy algorithm
\citep{zanardo13,laki-atca}. 

\renewcommand{\textfraction}{0.05}
\renewcommand{\topfraction}{0.95}
\renewcommand{\bottomfraction}{0.95}
\renewcommand{\floatpagefraction}{0.1}

\section{Results}

In Figure~1, we present the first spatially resolved submillimeter
continuum observations of SN$\,$1987A, obtained with the Atacama Large
Millimeter/Submillimeter Array (ALMA).  

\begin{figure}[h]
\centerline{\resizebox{6in}{!}{\includegraphics{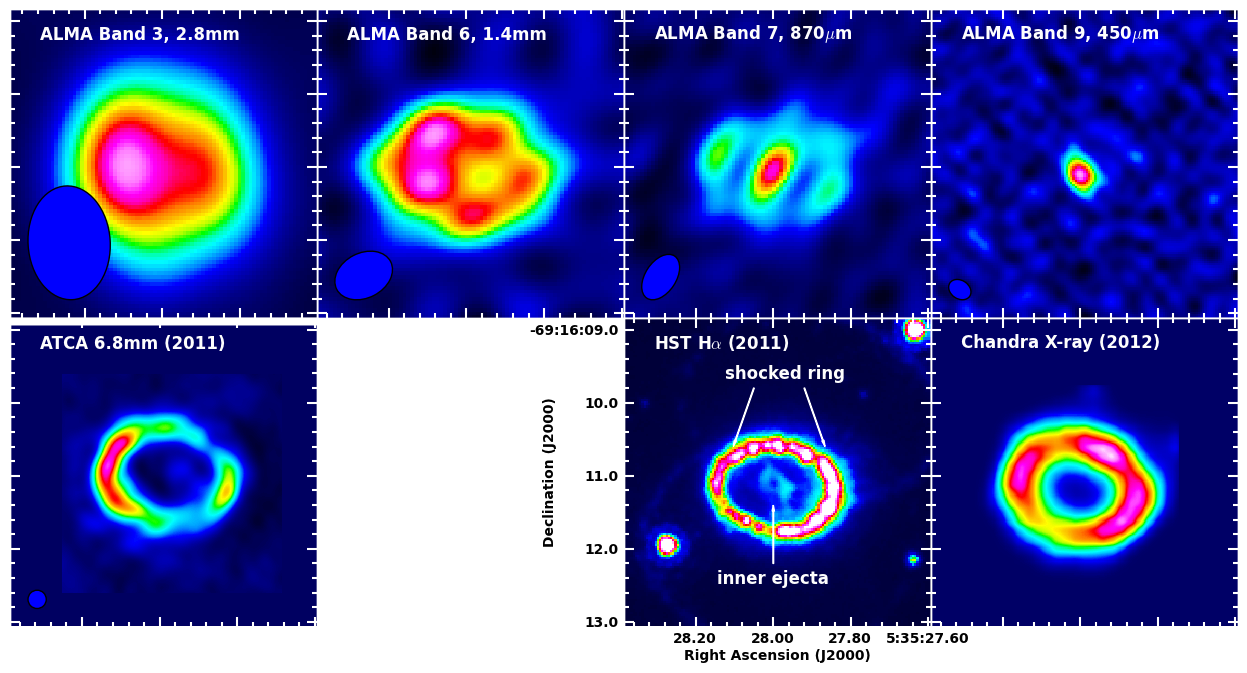}}}
\caption{ Top row: continuum images of SN$\,$1987A in ALMA bands 3, 6,
  7, and 9 (2.8$\,$mm, 1.4$\,$mm, 870$\,$\um and 450$\,$\um
  respectively).  The spatial resolution is marked by dark-blue
  ovals. In band~9 it is 0.33$\times$0.25\arcsec, 15\% of the diameter of the
  equatorial ring.  At bands 7, 6, and 3 the beams are 0.69$\times$0.42\arcsec, 0.83$\times$0.61\arcsec, and 1.56$\times$1.12\arcsec, respectively.
 At long wavelengths, the emission is a torus
  associated with the supernova shock wave; shorter wavelengths are
  dominated by the inner supernova ejecta.  The bottom row shows
  images of the continuum at 6.8$\,$mm imaged with the Australia Telescope
  Compact Array ATCA \citep[][0.25\arcsec\xspace beam]{zanardo13}, the Hydrogen H$\alpha$ line
  imaged with the Hubble Space Telescope HST \citep[Image courtesy of
  R. Kirshner and the SAINTS collaboration, see also ][]{larsson13},
  and the soft X-ray emission imaged with the Chandra X-ray
  Observatory \citep{helder13}.}
\end{figure}

We can now separate emission from the torus from that of the inner
ejecta at 450$\,$\um, 870$\,$\um, and 1.4$\,$mm.
Decomposition of the torus and inner ejecta was performed in the image
plane.  The torus was represented with the clean model (collection of
point sources or clean components) from deconvolving the ATCA
6.8$\,$mm image.  The inner ejecta were represented by the ALMA
Band~9 (450$\,$\um) model.  For each band, the image was decomposed
into a clean model and residual. The torus or ejecta model, scaled in
amplitude, was subtracted from the model, which was then restored with
the appropriate Gaussian restoring beam, and the residual image added
back in.  This method results in minimal noise from the torus or
ejecta observation being introduced, merely preserving the noise
already present in each original image.  The amplitude scaling is
varied to minimize the residuals in the subtracted image. The torus
and ejecta were also subtracted from each other in the Fourier plane,
with consistent results.  The torus and ejecta flux densities measured
using these independent decompositions were combined to determine the
flux densities and uncertainties listed in Table~\ref{fluxtable}.

\begin{deluxetable}{llllllll}
\renewcommand{\tabcolsep}{0.5ex}
\tablecaption{Flux densities\label{fluxtable}.}
\tablehead{ \colhead{Component} & \colhead{$\nu$} & \colhead{$\lambda$} & \colhead{F$_\nu$}& \colhead{Epoch} & \colhead{Telescope} & \colhead{Angular} & \colhead{Ref.}\\
&\colhead{[GHz]}&&\colhead{[mJy]}&&&\colhead{Res.}&}
\startdata
torus   & 36.2 & 8.3 mm  & 27$\pm$6     & 2008 & ATCA     & 0.3\arcsec  & \citet{potter09} \\
torus   & 44   & 6.8 mm  & 40$\pm$2     & 2011 & ATCA     & 0.3\arcsec  & \citet{zanardo13}\\
torus   & 90   & 3.2 mm  & 23.7$\pm$2.6 & 2011 & ATCA     & 0.7\arcsec  & \citet{laki-atca} \\
both    & 110  & 2.8 mm  & 27$\pm$3     & 2012 & ALMA     & 1.3\arcsec  & this \\
torus   & 215  & 1.4 mm  & 17$\pm$3     & 2012 & ALMA     & 0.7\arcsec  & this \\
ejecta  & 215  & 1.4 mm  & $<$2         & 2012 & ALMA     & 0.7\arcsec  & this \\
torus   & 345  & 870 \um & 10$\pm$1.5   & 2012 & ALMA     & 0.5\arcsec  & this \\
ejecta  & 345  & 870 \um & 5$\pm$1      & 2012 & ALMA     & 0.5\arcsec  & this \\
torus   & 680  & 440 \um & $<$7         & 2012 & ALMA     & 0.3\arcsec  & this \\
ejecta  & 680  & 440 \um & 50$\pm$15    & 2012 & ALMA     & 0.3\arcsec  & this \\
both    & 860  & 350 \um & 54$\pm$18    & 2010 & Herschel & 24\arcsec   & \citet{matsuura11}\\
both    & 860  & 350 \um & 44$\pm$7     & 2011 & APEX     & 8\arcsec    & \citet{laki-saboca}\\
both    & 1200 & 250 \um & 123$\pm$13   & 2010 & Herschel & 18\arcsec   & \citet{matsuura11}\\
both    & 1900 & 160 \um & 125$\pm$42   & 2010 & Herschel & 9.5\arcsec  & \citet{matsuura11}\\
both    & 3000 & 100 \um & 54$\pm$18    & 2010 & Herschel & 13.5\arcsec & \citet{matsuura11}\\
\enddata
\end{deluxetable}

\section{Discussion}

\begin{figure}[h]
\centerline{\resizebox{6in}{!}{\includegraphics{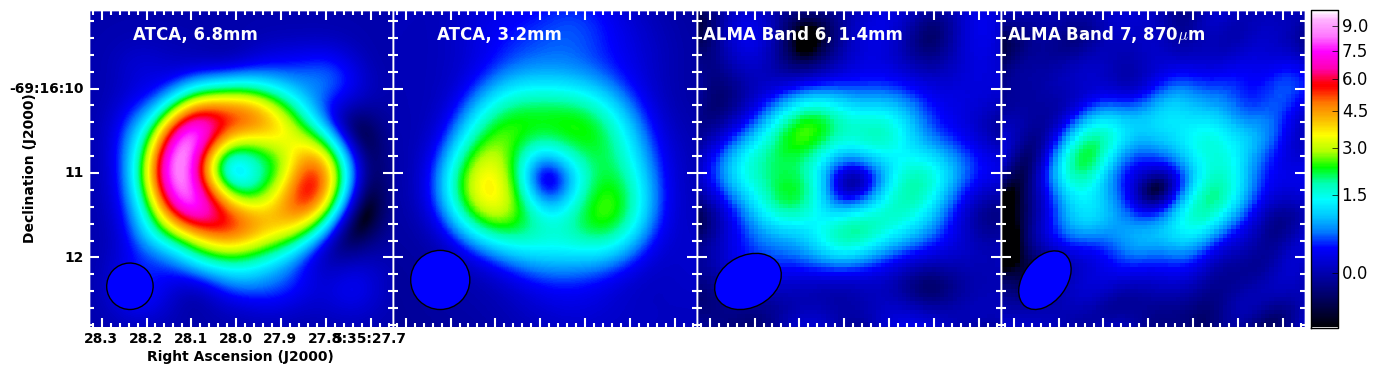}}}
\caption{\label{torusfig} Emission from the synchrotron torus of
  SN$\,$1987A as a function of wavelength, on the same intensity color
  scale. Emission from the inner ejecta has been subtracted to isolate
  the torus. The 6.8$\,$mm ATCA image \citep{zanardo13} has been
  smoothed to 0.55\arcsec\xspace beam similar to the ATCA 3.2$\,$mm 
  \citep[0.7\arcsec,][]{laki-atca} and ALMA images.}
\end{figure}

We discuss the torus first since the physical emission mechanism,
sychrotron emission from shock-accelerated particles, is less
controversial than the inner ejecta emission mechanism.  Images of the
torus alone (central ejecta removed) are shown in
Figure~\ref{torusfig}.  The eastern side is brighter at all (sub)millimeter
wavelengths, and the asymmetry decreases at shorter wavelengths.
In contrast, H$\alpha$ emission \citep{larsson13} from shocks being driven
into the equatorial ring, and soft X-ray emission \citep{helder13}
from hot plasma behind the shocks is now brighter in the west
(Fig.~1).  All wavelengths shown have brightened with time, and the
emission distribution has become geometrically flatter in the
equatorial plane, as the shocks interact with the dense ring
\citep{ng13,racusin09}.  Asymmetry results from differences in how far
that interaction has progressed \citep[e.g. the X-ray asymmetry can be
explained by faster shocks in the east;][]{zhekov09}.

\begin{figure}[h]
\centerline{\resizebox{6in}{!}{\includegraphics{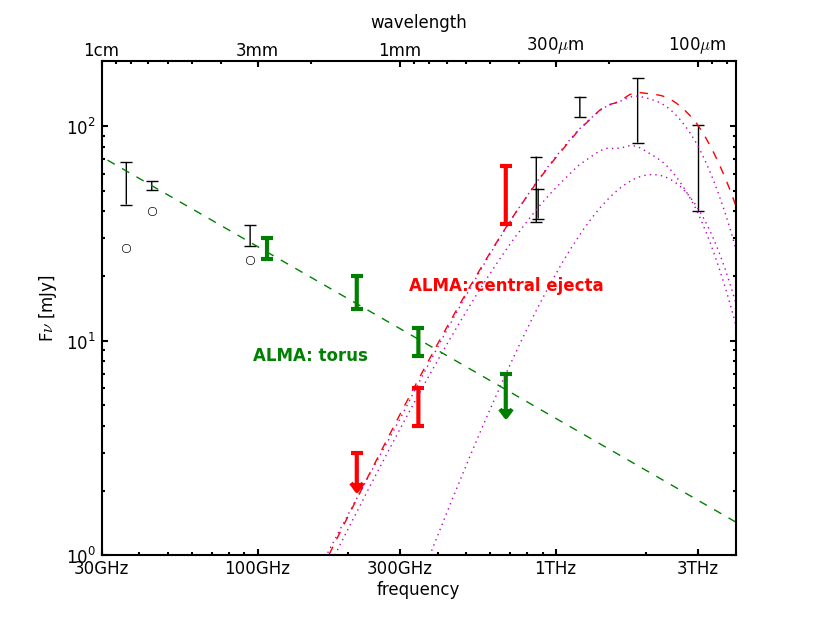}}}
\caption{\label{sedfig} Spatially separated ALMA flux densities of the
  torus (green) and inner ejecta (red).  Previous measurements are
  marked in black
  \citep{potter09,zanardo13,laki-atca,laki-saboca,matsuura11}. Measurements
  at longer wavelengths dominated by shock emission have been scaled
  to the epoch of the ALMA observations according to the light curve
  F$_\nu\propto$e$^{((t-5000)/2231)}$ at 44$\,$GHz \citep{zanardo10};
  the original flux densities at their epochs of observation are shown
  as open circles.  The spectral energy distribution (SED) of the
  torus is a power-law F$_\nu\propto\nu^\alpha$ with a single index
  $\alpha$ = -0.8$\pm$0.1 (green dashed line).  The SED of the inner
  ejecta is fit well by a model of dust emission -- shown here is
  0.23$\,$M$_\odot$ of amorphous carbon dust at 26$\,$K (red dashed
  line), and a combination of amorphous carbon and silicate dust
  (0.24$\,$M$_\odot$ and 0.39$\,$M$_\odot$ respectively, both at
  22$\,$K, two lower magenta dotted lines sum to the upper dotted
  line).}
\end{figure}

The integrated spectrum of the torus (Figure~\ref{sedfig}) is fit from
17$\,$mm to 870$\,$\um by a power-law with a spectral index $\alpha$ =
-0.8$\pm$0.1 (F$_\nu\propto\nu^\alpha$) and no evident spectral break.
This rules out significant synchrotron losses that would steepen the
spectrum, or contamination from free-free emission that would flatten
it.  The spectrum is quite steep (few high energy particles), compared
to what would be produced by diffusive shock acceleration in the test
particle limit, in a single shock at the observed $\sim$2000$\,$km$\,$s$^{-1}$
expansion velocity \citep{helder13,ng13}.  The shock structure is therefore
likely being modified by pressure from the accelerated particles
\citep{ellison00} or by an amplified tangled magnetic field
\citep{kirk96}.

The torus with its simple spectrum can now be removed from the images
to isolate the inner ejecta (Fig.~4).  Inner ejecta emission is well
resolved at 450$\,$\um, with a beam-deconvolved full-width at half-max
(FWHM) of 0.3$\pm$0.03\arcsec\xspace by 0.16$\pm$0.05\arcsec, or
2.2$\pm$0.2$\times$10$^{17}\,$cm by 1.2$\pm$0.4$\times$10$^{17}\,$cm,
corresponding to constant expansion velocities of 1350$\pm$150$\,$km$\,$s$^{-1}$
by 750$\pm$250$\,$km$\,$s$^{-1}$.  It is marginally resolved at 870$\,$\um;
constrained by the observations to have FWHM$<$0.4\arcsec, consistent
with the 450$\,$\um size.

\begin{figure}[h]
\centerline{\resizebox{6in}{!}{\includegraphics{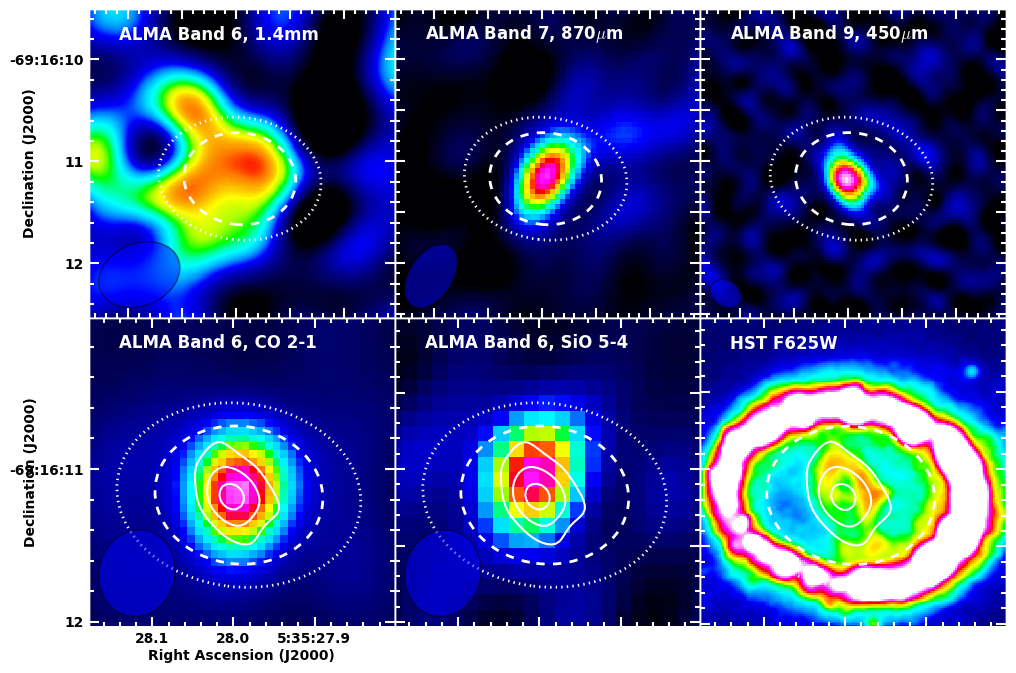}}}
\caption{\label{ejectafig}
  The inner ejecta of SN$\,$1987A.  The top row shows the ALMA images
  with the torus subtracted (location marked by dotted ellipse).  The
  bottom row shows the ALMA integrated images of emission from carbon
  monoxide and silicon oxide \citep[][0.57$\times$0.5\arcsec\xspace beam]{kam13}, and an HST F625W (optical)
  image (image courtesy of R. Kirshner and the SAINTS collaboration)
  with 450$\,$\um contours. The 450$\,$\um emission has the same
  north/south elongation as the optical and NIR \citep{larsson13}
  emission, and the 450$\,$\um peak may correspond to a hole in the
  optical emission.  Images are not at the same intensity scale; the
  dashed line is the location of the reverse shock
  \citep{michael03,france10}.}
\end{figure}

The spectrum of the inner ejecta rises steeply with frequency,
inconsistent with PWN emission, but very well modeled by dust.  The
ALMA and Herschel photometry can be fit with
0.23$\pm$0.05$\,$M$_\odot$ of amorphous carbon dust at 26$\pm$3$\,$K
(Fig.~3).  This implies that that the carbon dust mass is much higher
today than the 5$\times$10$^{-4}\,$M$_\odot$ measured two years after
the explosion \citep{wooden93,ercolano07}, and that within
uncertainties nearly all of the 0.23$\pm$0.1$\,$M$_\odot$ of carbon
\citep{thielemann90,woosley07} released in the explosion is now in
dust. 
The dust mass depends on opacity, temperature, and optical depth but
is quite robustly constrained by the data.  We use amorphous carbon
opacity $\kappa$=10 and 3$\,$cm$^2$g$^{-1}$ at 450 and 870$\,$\um,
respectively \citep{rouleau91}.  Values in the literature range from
2-10$\,$cm$^2$g$^{-1}$ at 450$\,$\um
\citep[e.g.][]{planck13,jager98}; using a lower value would
only raise the dust mass.
Our analysis of CO emission observed with ALMA and {\em Herschel}
finds $>$0.01$\;$M$_\odot$ of CO in SN1987A \citep{kam13}.  Within
uncertainties, the carbon in dust and CO doesn't yet exceed the
nucleosynthetic yield, but as future observations refine the CO and
dust masses, strong constraints may be placed on nucleosynthesis,
chemistry, dust coagulation, or all three.

The temperature is quite well constrained by these data, and theoretical models predict a similar temperature. Cooling by adiabatic expansion and radiation is offset by heating from $^{44}$Ti decay and by external X-ray heating from the shocks.  The current gas temperature in the absence of X-ray heating is modeled to be 20-100$\,$K \citep{fransson13}.  Models predict that significant X-rays do not yet penetrate inwards as far as the 2$\times$10$^{17}\,$cm radius of the ALMA 450$\,$\um emission \citep{fransson13}.  The compact and centrally peaked 450$\,$\um emission supports this interpretation, since external heating would likely result in more limb-brightened or extended dust emission.  Since the ionization fraction is below 1\% \citep{larsson13}, less than $\sim$40\% of the X-ray flux will go into heating \citep{xu91}, and $<$5\% of the observed total flux 4.7$\times$10$^{36}\,$erg$\,$s$^{-1}$ \citep{helder13} is intercepted by the ejecta.  Even if it did reach the dusty inner core, the energy deposition would be $<$10\% of the heating from $^{44}$Ti decay \citep{jerkstrand11}.   The best-fit dust mass M$_d$ scales approximately as T$_d^{-2}$, and M$_d >$0.1$\,$M$_\odot$ for T$_d <$50$\,$K.

The emission is optically thin at 450$\,$\um, and M$_d$ is insensitive
to unresolved clumpiness: If 0.23$\,$M$_\odot$ of dust uniformly
filled a region the size of the ALMA emission, the peak surface
density would be 0.02$\,$g$\,$cm$^{-2}$, with an optical depth
$\tau_{450} < $0.2.  If the dust is clumpy, each clump remains
optically thin unless the filling fraction is less than $\sim$2\%; a
filling fraction of 10-20\% was fitted to the infrared spectrum
\citep{lucy91,ercolano07} and consistent with the CO clump filling
factor of 0.14 fitted to ALMA CO emission \citep{kam13}.  Analytical
formulae for radiative transfer in dense clumps
\citep{varosi99} indicate that the effective optical depth of the
ensemble of clumps remains low over a very wide range of clump filling
fraction and number.

The data can also be fit by a combination of carbonaceous and silicate
(Mg$_2$SiO$_4$) dust (Fig.~3; dotted line), although the mass of
carbonaceous dust cannot be significantly reduced because amorphous
carbon has the highest submillimeter opacity $\kappa$ among
minerologies similar to the interstellar medium, and eliminating it
would require dust masses of other compositions larger than the
available metal mass (the best fit using pure silicate is
4$\,$M$_\odot$ at 21$\,$K).  Although these data do not strongly
constrain the mass of silicate dust, it is unlikely that only carbon
condensed in the SN$\,$1987A, given the presence of large amounts of
Mg and Si in the ejecta (0.065$\pm$0.1 and 0.19$\pm$0.1$\,$M$_\odot$
respectively) and the detection of SiO molecules at early epochs
\citep{roche91,wooden93}.  The absence of the 10$\,$\um feature during
days 615-775 puts a strong upper limit (15\%) on the fraction of dust
which is silicates, but does not exclude the presence of some silicate
dust during that epoch, since the ejecta could have been optically
thick at that wavelength \citep{ercolano07}. The silicate mass could
be much larger now, similar to the much larger carbon mass.

Models and observations show that core collapse supernovae like
SN$\,$1987A are highly inhomogeneous, with instabilities and
radioactive energy deposition mixing the initially chemically
stratified stellar interior into clumps with different compositions,
macroscopically mixed throughout the ejecta volume
\citep{mccray93,jerkstrand11}.  The distribution of
emission from different atoms, molecules, and dust not only reveals
the chemistry of their formation, but a snapshot of the early
supernova interior.  The extent of dust emission corresponds well to
line emission from CO$\,$2-1 and SiO$\,$5-4 observed simultaneously
with ALMA \citep{kam13}.  The 1-2$\times$10$^{17}\,$cm radial extent
corresponds to expansion velocities of 750-1400$\,$km$\,$s$^{-1}$,
consistent with the 1250$\,$km$\,$s$^{-1}$ half-width of the CO
emission \citep{kam13}, and within the commonly used maximum core
expansion velocity of $\sim$2000$\,$km$\,$s$^{-1}$
\citep{jerkstrand11}.  SiO is an important precursor to forming dust,
so the relative distributions of dust and SiO in SN$\,$1987A that ALMA
will be able to measure with higher spatial resolution in the future
will constrain formation theories.

In the last few years, the inner ejecta have developed a complex
morphology in optical and infrared emission, including north-south
elongation and a ``hole'' or fainter region in the center. The
450$\,$\um emission peak is coincident with the hole, and although the
optical emission is likely limb-brightened due to external X-ray
heating \citep{larsson11}, the dust we detect does have significant
optical opacity.  Smooth dust would have $\tau_V>$1000, and 100 dusty
clumps filling 10\% of the inner ejecta volume would have
$\tau_V\sim$1.7 \citep{varosi99}. (About 100 clumps were predicted by
collapse simulations \citep{hammer10} and fitted to observed optical
lines \citep{jerkstrand11}).

\section{Conclusions}

We have used the powerful resolution of ALMA to show clearly that
prodigious amounts of dust have formed in the cold inner ejecta of
SN$\,$1987A.  Our data suggests that nearly all of the carbon has
condensed into dust, so condensation must be efficient.  A large mass
of clumpy dust was tentatively detected at 1.3$\,$mm 4 years after the
explosion \citep{biermann92} -- if this is the same dust then it
formed quite rapidly.  If dust production in other supernovae
resembles that in SN$\,$1987A, then core-collapse supernovae might
contribute as much dust to galaxies as asymptotic giant branch stars.
In the absence of grain destruction, dust in high-redshift galaxies
can be explained with only 0.1$\,$M$_\odot$ produced per type II SN
\citep{dwek07}. SN$\,$1987A has the largest measured dust mass, but a
few other remnants contain almost 0.1$\,$M$_\odot$ (Cassiopeia A,
\citet{barlow10}; Crab Nebula, \citet{gomez12}).  However, dust formed
in a supernova must survive passage through the reverse shock to be
dispersed into the interstellar medium, and then also survive shock
passages once dispersed \citep{jones11}.  The reverse shock in
SN$\,$1987A located with spatially resolved H$\alpha$ spectra
\citep{michael03,france10} has a minimum radius in the equatorial
plane of $\sim$4$\times$10$^{17}\,$cm, significantly larger than the
450$\,$\um emission (Fig.~4).  The new dust has not been significantly
processed by the reverse shock.  Models of dust destruction predict a
wide range of survival fractions \citep{nozawa07}.  If
0.23$\,$M$_\odot$ of dust is typically created in type II supernovae,
and 0.1$\,$M$_\odot$ passes through the reverse shock into the ISM,
then SN$\,$II could dominate dust production in galaxies at all
redshifts.

\section{Acknowledgements}
This paper makes use of the following ALMA data: ADS/JAO.ALMA\#2011.0.00273.S (PI Indebetouw). ALMA is a partnership of ESO (representing its member states), NSF (USA) and NINS (Japan), together with NRC (Canada) and NSC and ASIAA (Taiwan), in cooperation with the Republic of Chile. The Joint ALMA Observatory is operated by ESO, AUI/NRAO and NAOJ.
The National Radio Astronomy Observatory is a facility of the National Science Foundation operated under cooperative agreement by Associated Universities, Inc.
M. Meixner was supported by NASA NAG5-12595 and NASA/ADAP NNX13AE36G.

\end{document}